# Source REconstruction of Arcs behind Cluster Halos (SourceREACH): A New Source Reconstruction Algorithm Optimized for Giant Arcs and Galaxy Cluster Lenses

Lana Eid and Charles R. Keeton
Rutgers University–New Brunswick, Department of Physics and Astronomy, 136 Frelinghuysen Rd., Piscataway, NJ 08854, USA
*Received 2025 November 14; revised 2026 April 15; accepted 2026 April 16; published 2026 May 21*

## Abstract

We introduce a new algorithm designed for reconstructing extended lensed images, specifically giant arcs lensed by galaxy clusters. These highly magnified images contain spatially unique information about both the most uncertain regions of the cluster mass distribution as well as the properties of the background source, but modeling them requires significant computational effort. Our new source reconstruction methodology is designed to be accurate and efficient for high-resolution observations in which point-spread function (PSF) effects are not significant. The overall process deconvolves the observed image by the PSF, delenses the image pixels, and uses interpolation or regression with smoothing to determine the model source. By working with delensed points, the method accounts for varying resolution across the source plane while avoiding the need to construct and manipulate large matrices. We evaluate the speed and accuracy of different interpolation and regression methods using both mock data and real data for the giant arc in Abell 370. We find that utilizing k-nearest neighbors regression results in the best balance of noise smoothing and preservation of compact detail in the source.

## 1. Introduction

Galaxy cluster lens models inform us about the cluster-scale mass distribution (e.g., J. Richard et al. 2010; C. A. Raney et al. 2020a; J. M. Diego et al. 2023b; M. Limousin et al. 2025), possible substructure on a smaller scale (e.g., J. M. Diego et al. 2024), and cosmological parameters like the Hubble constant and dark energy from time delays (e.g., J. Vega-Ferrero et al. 2019; G. B. Caminha et al. 2022; G. Granata et al. 2025; J. D. R. Pierel et al. 2026). Cluster lens models are often constrained with collections of compact images; each cluster can have multiple compact lensed sources, which can lead to dozens of images that provide constraints spanning the cluster field (e.g., C. E. Sharon et al. 2016; J. M. Lotz et al. 2017).

Some lensed sources are extended, leading to resolved images that offer further, unique constraints. Source reconstruction models facilitate spatially resolved studies of the sources, including properties such as kinematics (e.g., V. Patrício et al. 2018), star formation rate at the scale of tens of kiloparsecs (e.g., T. L. Johnson et al. 2017b; V. Patrício et al. 2018), and locations of emission line regions such as Lyman $\alpha$ (e.g., A. Claeyssens et al. 2022; A. Navarre et al. 2024). Giant, spatially extended arcs are highly magnified, multiply lensed extended objects that cross lensing critical curves, defined as the theoretically infinite regions of magnification predicted by a mass model for a given lens and source redshift pair. Consequently, they offer maximum insight into both the source in the early Universe as well as the lensing regions producing these images.

Successful modeling of giant lensed arcs must handle both the mass model and the source simultaneously. Techniques for such combined analyses are often used for galaxy-scale (e.g., S. Birrer & A. Amara 2018; A. Galan et al. 2022; L. Biggio et al. 2023; J. Rüstig et al. 2024) and even group-scale systems (e.g., H. Wang et al. 2022; H. Ding et al. 2025) but are computationally expensive for cluster-scale models. Various approaches have been taken in the literature to forward the cluster modeling part of the field using separate, multiply lensed, extended images (e.g., J. M. Diego et al. 2016; L. Yang et al. 2020; A. Acebron et al. 2024) but without handling full, extended giant arcs specifically.

In-depth studies, including of high-redshift source properties (T. L. Johnson et al. 2017b; C. E. Sharon et al. 2019; A. Claeyssens et al. 2022; J. M. Diego et al. 2023a; Y. Fudamoto et al. 2025; A. Navarre et al. 2024), small-scale lensing dark matter structure effects of milli-lensing (e.g., A. Niemiec et al. 2023; J. M. Diego et al. 2025; Y. Fudamoto et al. 2025; M. Limousin et al. 2025), and highly magnified microlensing events (e.g., J. M. Diego et al. 2024, 2025) show that it is particularly important to have accurate models near cluster critical curves. Yet results from the Hubble Frontier Fields (HFF; J. M. Lotz et al. 2017) multi-team modeling efforts (e.g., C. A. Raney et al. 2020b; L. Yang et al. 2020; L. Eid & C. R. Keeton 2025) and from modeling code comparisons on recent JWST data (e.g., A. Agrawal et al. 2026) have demonstrated that these regions are still underconstrained, with wide variation and systematic uncertainty even among models that generally fit lensed point images equally well (C. A. Raney et al. 2020b; A. Galan et al. 2023, 2024a, 2024b; K. Lasko et al. 2023; J. Liesenborgs et al. 2024; M. Limousin et al. 2025).

Many studies have examined how lensed arcs constrain local regions of cluster mass models (e.g., S. Birrer 2021; J. Lin et al. 2023; A. Ç. Şengül et al. 2023; A. Ç. Şengül 2025). However, care must be taken to ensure that a mass model locally optimized for a giant arc does not lose validity for the entire cluster due to the focus on more accurately constraining just that region (e.g., A. Ç. Şengül 2025). J. Lin et al. (2023) show that matching different regions across extended images can adjust local regions of a cluster model without affecting the global properties of the cluster model. We similarly showed this for the case of the giant arc in Abell 370 (A370; L. Eid & C. R. Keeton 2025, hereafter Paper I), in which we locally optimized a fiducial cluster model from C. A. Raney et al. (2020a) using a full pixel-based source reconstruction without significantly affecting the rest of the lensed image position fits or adjusting the cluster-scale matter halos.

It is important to develop a comprehensive approach to modeling cluster lenses that combines a global analysis of compact image collections with a detailed, local analysis of caustic-crossing giant arcs. This work, which is the second in a series of three papers, focuses on developing and testing a source reconstruction methodology accurate and efficient enough to use as a cluster model constraint valid for the entire field; this is particularly important in the context of the wealth of information from present and upcoming targeted observations and large surveys (A. J. Shajib et al. 2025).

Source reconstruction can be done with various techniques, including forward-modeling with basis functions and iteratively added complexity (e.g., A. Refregier 2003; L. Yang et al. 2020; A. Ç. Şengül et al. 2023); semi-linear inversion (e.g., S. J. Warren & S. Dye 2003; J. W. Nightingale & S. Dye 2015; R. Joseph et al. 2019) and pixel-based (e.g., S. H. Suyu et al. 2006; S. Vegetti & L. V. E. Koopmans 2009; A. S. Tagore & C. R. Keeton 2014; A. J. Young et al. 2022) techniques that trade computational speed to better capture source detail; ray-tracing (e.g., J. M. Diego et al. 2016; N. Tessore et al. 2016) and hybrid (e.g., J. Rüstig et al. 2024) methods; backprojection using a modeled magnification matrix (e.g., R. E. Griffiths et al. 2021); and, more recently, machine learning-guided methods (e.g., W. R. Morningstar et al. 2019; K. Karchev et al. 2022).

Bayesian probabilistic methods are often built into these methods to incorporate uncertainties and regularization; varying source plane resolution can be handled using adaptive gridding techniques, including Delaunay triangulation or Voronoi tessellation, and weighting based on pixel flux. Many of these methods rely on grids in the image and source planes, which are connected through large matrices. Manipulating those matrices requires significant computational resources in terms of both memory and time; fully modeling long, critical curve-crossing arcs can take weeks or months (e.g., A. Acebron et al. 2024).

For example, in Paper I, we used the pixel-based source reconstruction code `pixsrc`, which is based on the Bayesian formalism from S. H. Suyu et al. (2006), with adaptive gridding following S. Vegetti & L. V. E. Koopmans (2009). Our local optimization had ∼100,000 pixels within the data mask, and we varied two parameters for each of the 11 local cluster masses. The full cluster field mass model included 276 mass components with several parameters each, which was not practically feasible to optimize with the standard grid-based Bayesian framework.

Here, we introduce a new source reconstruction methodology, Source REconstruction of Arcs behind Cluster Halos (`SourceREACH`), which is intended to facilitate using giant arcs as constraints in galaxy cluster mass modeling using high-resolution data with minimal point-spread function (PSF) effects. The methodology is designed to work with any set of standard output maps, such as from the HFF lens modeling effort (J. M. Lotz et al. 2017; L. Yang et al. 2020). Our approach avoids storing and manipulating large matrices, reducing the resource demands and yielding faster run times. The key idea is to apply interpolation or regression methods with smoothing to the delensed data points. We perform a comprehensive study comparing different adaptive smoothing methods from big data analysis. We then validate our final adaptive smoothing choice using HFF optical data (J. M. Lotz et al. 2017) and compare it to our previous source reconstruction for A370 in Paper I. Once we ensure this accuracy, we intend to use this methodology to fold in the giant arc source reconstruction as a constraint in the global A370 cluster model in Paper III of this series (L. Eid & C. R. Keeton 2026, in preparation).

Section 2 presents the data, mock data, and our previously optimized model; Section 3 presents the methodological development and testing; Section 4 presents our results; Section 5 interprets the results; and Section 6 contextualizes them for future studies.

## 2. Data and Models

The `SourceREACH` methodology requires an isolated surface brightness map of a lensed arc. In this section, we describe the process for obtaining such a map from HFF observations of A370, which involves removing cluster member galaxies and deconvolving with the PSF. We also describe mock data generated for testing purposes. source and image plane parameters for our purposes here.

Throughout this paper, we hold the mass model fixed, making use of the 2D cluster mass model from C. A. Raney et al. (2020a). Utilizing these data and models will allow us to directly compare the methodology presented here with the established pixel-based source reconstruction code that we used in Paper I.

### *2.1. The A370 Galaxy Cluster Field*

We use data from the HFF program (J. M. Lotz et al. 2017), which was a 3 yr, 840-orbit Hubble Space Telescope (HST) targeting six galaxy cluster fields with the aim of obtaining deep images of the clusters and faint $z \sim 5$–10 background galaxies magnified by strong lensing. This survey program is ideal for our study, since it is the base of the various comparison studies that have followed from the lens modeling teams that used the same data to construct their cluster models (e.g., C. A. Raney et al. 2020b; L. Yang et al. 2020).

We focus on the $z = 0.375$ A370 cluster field, which includes the first confirmed strongly lensed giant arc, from a galaxy at $z = 0.725$ (F. Hammer 1987; G. Soucail 1987; G. Soucail et al. 1987a, 1987b, 1988). We chose this well-studied system of an extended, critical curve-crossing, giant arc comprised of five merging images of the same background galaxy for our in-depth study of how a high-resolution source reconstruction can be efficiently leveraged as a constraint in order to help address systematic differences between results from diverse cluster modeling methods that persist in these predicted high-magnification regions of interest in these cluster models (C. A. Raney et al. 2020b; A. Galan et al. 2024b).

Paper I details the steps we took to isolate the arc and format our inputs for each filter and corresponding data image; we briefly outline them here. We use the highest HFF signal-to-noise filter, F814W, to develop our methodology; we then apply this methodology to the HFF F814W, F606W, and F435W filters for comparison to our red, green, blue (RGB) source reconstruction results in Paper I. The top left panel of Figure 1 shows the data image from the F814W filter, which we used in Paper I to optimize the cluster model. In order to isolate the brightness pixels of the giant arc so we can trace them back, we used the `isophote` package from `python-photutils` (L. Bradley et al. 2022) to model and remove the contaminating light from cluster members and other galaxies projected close to the giant arc (upper right panel, Figure 1). We measured the noise in an isolated region away from the giant arc as $\sigma = 0.00038$ in the fiducial units of electrons/second. Our data mask was created directly in `SAOImageDS9` (W. A. Joye & E. Mandel 2003) using the cleaned, isolated arc image plus a small margin of background, excluding areas with any significant artifacts (lower left panel, Figure 1). For each filter and corresponding data image, we used the `EPSFBuilder` function from the `pythonphotutils` (L. Bradley et al. 2022) package to create an autonormalized "effective PSF" for the filter from an input set of isolated stars in the filter image, which we used to deconvolve the image data from the PSF using Richardson–Lucy deconvolution (lower right panel, Figure 1) (W. H. Richardson 1972; L. B. Lucy 1974).

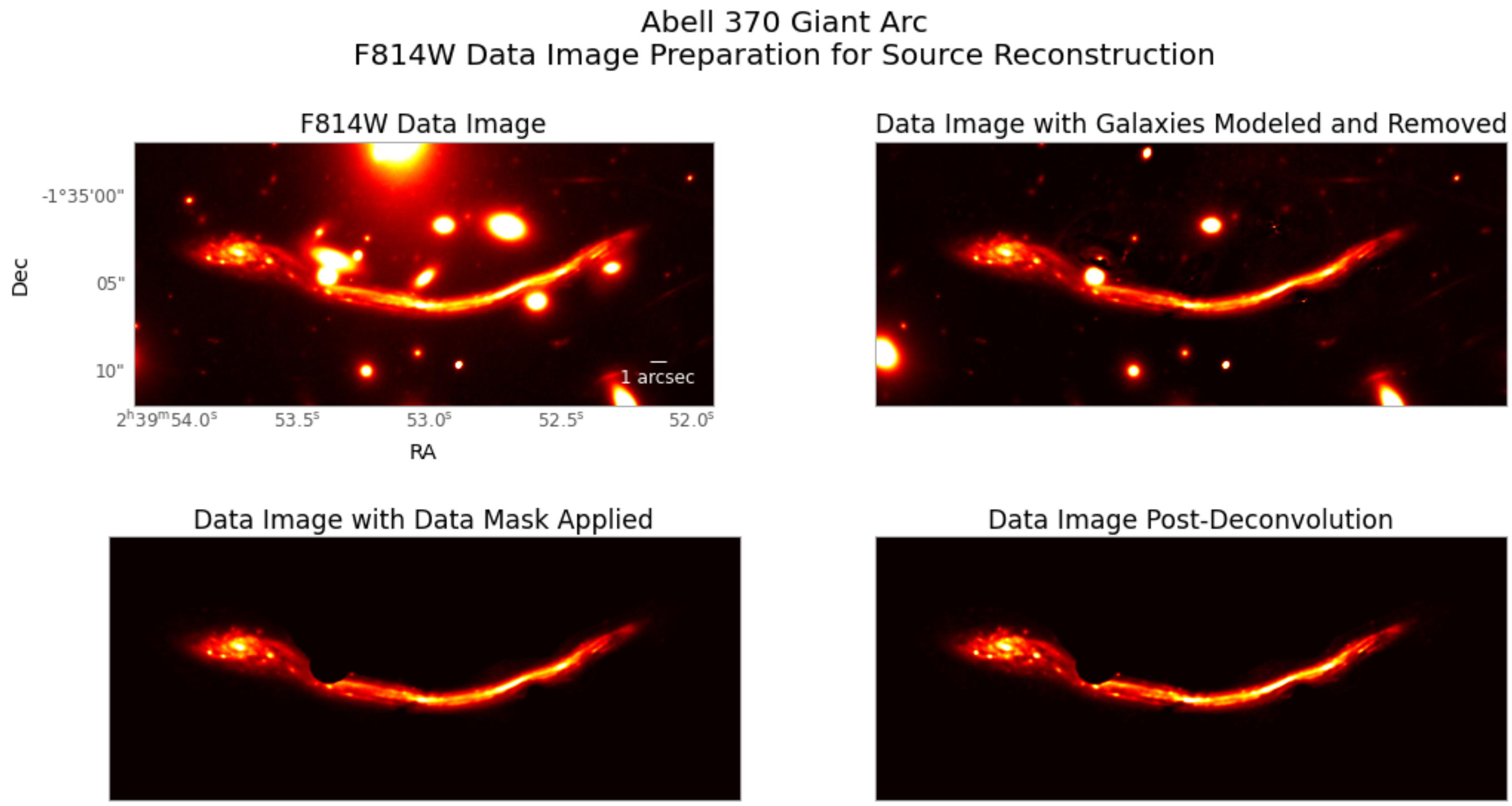

**Figure 1.** Top left: data image created from the HST F814W filter from the HFF program. Top right: data arc with light from galaxies near the arc modeled and subtracted (except for two along the left portion of the arc, which were masked instead; see Paper I for further discussion). In this map, the noise level is measured away from the arc to be $\sigma = 0.00038$ in the fiducial units of electrons per second. Bottom left: masked and cleaned data arc, which was the input to `pixsrc` for our previous optimization runs. Bottom right: cleaned and masked data arc after deconvolution with the Richardson–Lucy algorithm.

### 2.2. Cluster Lens Model

Multiple lens modeling teams produced publicly available cluster models for the HFF through parametric (E. Jullo et al. 2007; M. Oguri 2010; M. Jauzac et al. 2012, 2014; T. L. Johnson et al. 2014; J. Richard et al. 2014; M. Ishigaki et al. 2015; R. Kawamata et al. 2016, 2018; G. B. Caminha et al. 2017; C. A. Raney et al. 2020a), nonparametric (M. Bradač et al. 2005, 2009; J. Liesenborgs et al. 2007), and hybrid (J. M. Diego et al. 2005a, 2005b, 2007, 2015; J. Vega-Ferrero et al. 2019) methods with a standard format and calibration.[1]

In Paper I, we began with the 2D fiducial lens model from C. A. Raney et al. (2020a), C. A. Raney et al. (2020b). That parametric model included 275 mass components: 256 cluster member galaxies, four cluster-scale matter halos representing contributions from both the large dark matter components as well as X-ray gas, and 15 line-of-sight galaxies. It was originally fit to 91 lensed images from 31 image systems in the cluster field. We optimized the model to better fit the giant arc by adjusting the properties of 11 galaxies projected near the arc, two of which were previously listed as a single galaxy prior to higher-resolution observations. That optimization was done using the pixel-based source reconstruction code `pixsrc` (A. S. Tagore & C. R. Keeton 2014; A. S. Tagore & N. Jackson 2016), which is based on the S. H. Suyu et al. (2006) source reconstruction Bayesian formalism.

We only used the F814W filter image for the local optimization in Paper I, as it had the most structure to fit and the highest signal-to-noise value. Consequently, the fits in the other two filters both improved alongside the red filter, but there is still room for improvement, particularly in the more concentrated, small-scale brightness regions similar to those commonly used as point-image constraints in other source reconstruction (e.g., T. L. Johnson et al. 2017a, 2017b).

We note that our cluster lens model remains fixed throughout this paper; no further optimization is done here, as our primary focus is ensuring our source reconstruction code is efficient and accurate as compared to the data and previous source reconstruction methods. We will use our source reconstruction code to further optimize the model in a follow-up study.

[1] https://archive.stsci.edu/prepds/frontier/lensmodels/

### 2.3. Mock Data

In addition to using the real data for A370, we also construct two mock arcs to serve as test cases for which we know the "truth." This allows us to determine whether our algorithm provides accurate results and understand how different approaches handle smoothing and noise.

The first mock arc is created by lensing the best source model from the `pixsrc` analysis, which produced a regularized and smoothed source. We work with an idealized map with no PSF blurring or added noise in order to verify that our delensing source reconstruction is accurate in the context of lensing.

We then created a version of our mock arc with Gaussian noise added in the image plane. For this test, we chose to inflate the noise relative to the measured value in order to investigate how each adaptive smoothing routine handles noise and whether it runs the risk of overfitting the noise while attempting to fit the signal. In particular, we chose a standard deviation of 0.0045 to match the noise level that was used in Paper I to achieve the desired level of smoothing in the previous `pixsrc` analysis. This corresponds to a peak signal-to-noise ratio of approximately 18.

## 3. SourceREACH Methodology

### 3.1. Source Reconstruction Algorithm

Our source reconstruction algorithm, which we construct in Python for portability, is a straightforward delensing code for high-resolution datasets in which atmospheric and optical blurring can be removed through PSF deconvolution. In general, if $\alpha_x(x, y)$ and $\alpha_y(x, y)$ are the $x$ and $y$ deflections, respectively, $S(u, v)$ is the surface brightness distribution in the source plane, and $I(x, y)$ is the surface brightness distribution in the image plane, we can make use of the conservation of surface brightness in lensing to relate the brightness in the image to the corresponding brightness in the source

$$I(x, y) = S(x - \alpha_x(x, y), y - \alpha_y(x, y)) \tag{1}$$

and to map each pixel back to a unique position in the source plane

$$u = x - \alpha_x(x, y), \tag{2}$$

$$v = y - \alpha_y(x, y). \tag{3}$$

The main inputs are an array of image pixel positions $(x, y)$ and an array of their original data brightnesses $B_{\rm obs}(x, y)$. The general workflow of the code is as follows:

1. Deconvolve the image $B_{\rm obs}(x, y)$ from the PSF using Richardson–Lucy deconvolution (W. H. Richardson 1972; L. B. Lucy 1974) to obtain a brightness array $B_{\rm data}(x, y)$.
2. Delens the deconvolved image $B_{\rm data}(x, y)$ by mapping all of the image plane brightness pixels back to the source plane using the input deflection maps. The resulting set of brightness points $B_{\rm data}(u, v)$ will have irregular spacing. In order to preserve the varying resolution of the source from different parts of the giant arc and not introduce any extra systematics, we work solely with this set of points without any additional gridding.
3. Perform the desired adaptive smoothing routine in the source plane, whether (a) an interpolation with smoothing or (b) a function trained on the data values that then predicts values at those same positions, to obtain an array of smoothed brightness values $B_{\rm smooth}(u, v)$. We explore four different adaptive smoothing routines in Section 3.2.
4. Relens the new brightness values in the source plane to their deflected image plane positions to obtain $B_{\rm smooth}(x, y)$, and convolve with the PSF to obtain the model image $B_{{\rm model}l}(x, y)$. A single rms or $\chi^2$ value from the image plane residuals $B_{\rm obs}(x, y) - B_{\rm model}(x, y)$ can then be used as a constraint for optimizing a cluster model.

In order to evaluate the accuracy and efficiency of each smoothing and prediction method, we utilize the rms value of the residuals between the data and model, computed in both the image plane and source plane. For the image plane, we only compute the rms within the data mask.[2] We calculate the image plane residual rms value as

$$\mathrm{rms} = \sqrt{\frac{1}{N}\sum_{i=1}^{N}(B_{{\rm obs},i} - B_{{\rm model},i})^2}, \tag{4}$$

where $N$ represents the number of data pixels in the data mask, which is 104,176 pixels in our case. We can also calculate the image residual $\chi^2$ value,

$$\chi^2 = \sum_{i=1}^{N}\frac{(B_{{\rm obs},i} - B_{{\rm model},i})^2}{\sigma^2}, \tag{5}$$

where $\sigma$ is the noise level calculated from the standard deviation of the background of the cluster field: $\sigma_{\rm F814W} = 0.00038$, $\sigma_{\rm F814W} = 0.00090$, $\sigma_{\rm F814W} = 0.00043$. We then use $\chi^2/N$ to assess the fit quality: a value $\chi^2/N \gg 1$ indicates a poor fit to the data, while a value $\chi^2/N \ll 1$ would suggest that the model is overfitting noise in the data. We report both rms and $\chi^2/N$ for completeness; the relation between the two quantities is

$$\frac{\chi^2}{N} = \frac{(\mathrm{rms})^2}{\sigma^2}. \tag{6}$$

### 3.2. Smoothing Methods

A critical part of source reconstruction is handling noise; this often involves some form of smoothing or regularization to avoid overfitting. G. Vernardos & L. V. E. Koopmans (2022) examine different approaches to regularization, including physically motivated priors, in order to break degeneracies in underconstrained lens modeling and source reconstruction. This works to avoid overfitting and ensure that differential magnifications in different regions of the source are handled properly (A. S. Tagore & C. R. Keeton 2014; A. S. Tagore &

[2] In Paper I, in order to accommodate the different model arc extents from different HFF teams, we previously calculated the rms within a slightly larger rectangular region with extra zero values. Here, we work strictly within the data mask valid for the giant arc data and our previously optimized model plus a margin of background with no extra zero values. The residual rms for the F814W filter source reconstruction was $3.53 \times 10^{-4}$ in Paper I; strictly within the data mask, we now calculate it as $2.19 \times 10^{-3}$.

N. Jackson 2016; G. Vernardos & L. V. E. Koopmans 2022; A. J. Young et al. 2022). Regularization may overly smooth these features if not done carefully because it avoids large changes when aiming for a smooth source. Compact regions, such as star-forming knots, are sometimes explicitly used in lens modeling, such as in the previous analyses of A370 by J. Richard et al. (2010) and V. Patrício et al. (2018), though their source reconstruction only utilized the least magnified, least warped, leftmost arc image.

Our approach uses interpolation or regression functions in the source plane to provide adaptive smoothing. We choose to test various techniques originally developed for big data analysis in order to maximize efficiency on the now purely computational question of how to constrain a two-dimensional function for the source from the set of scattered, noisy, delensed source points. We examine the following methods: radial basis function (RBF) interpolation, k-nearest neighbors (KNN) regression, decision tree regression (DTR), and random forest regression (RFR). Here, we compare these four interpolation routines with various inputs relating to smoothing for the real data in the HFF F814W filter. We select options for each smoothing method that avoid overfitting the delensed noise and that run on the order of a few seconds. For each method, we test different inputs and options, including parameters such as the number of nearest neighbors, number of estimators, and minimum leaf size, in order to best balance smoothing noise in the source with ensuring that the source is complete; we describe these below and illustrate the process in the Appendix.

**Table 1**
Comparison of Adaptive Smoothing Techniques and Hyperparameter Testing, All Applied to Our Isolated F814W Giant Arc Data

| RBF Interpolator (`neighbors = 6`, `smoothing = 0.25`) | rms | $\chi$2/N | Runtime (s) |
|---|---|---|---|
| `kernel=thin_plate_spline` | $1.86 \times 10^{-3}$ | 24.41 | 37.34 |
| Linear | $1.79 \times 10^{-3}$ | 22.74 | 36.41 |
| Cubic | $1.92 \times 10^{-3}$ | 26.08 | 35.89 |
| **KNN Regressor (`weight = uniform`)** | **rms** | **$\chi^2/N$** | **Runtime (s)** |
| Ball tree (`n_neighbors=6`) | $2.21 \times 10^{-3}$ | 34.54 | 32.01 |
| Ball tree (`n_neighbors=18`) | $2.26 \times 10^{-3}$ | 36.08 | 32.91 |
| KD tree (`n_neighbors=6`) | $2.21 \times 10^{-3}$ | 34.54 | 30.95 |
| KD tree (`n_neighbors=18`) | $2.26 \times 10^{-3}$ | 36.08 | 32.41 |
| **KNN Regressor (Ball tree, `weight = uniform`)** | **rms** | **$\chi^2/N$** | **Runtime (s)** |
| (`n_neighbors=6`) | $2.21 \times 10^{-3}$ | 34.54 | 32.00 |
| (`n_neighbors=15`) | $2.24 \times 10^{-3}$ | 35.44 | 32.42 |
| (`n_neighbors=25`) | $2.28 \times 10^{-3}$ | 36.64 | 33.07 |
| (`n_neighbors=50`) | $2.30 \times 10^{-3}$ | 37.45 | 34.19 |
| **Decision Tree Regressor (`min_samples_leaf=6`, `random_state=42`)** | **rms** | **$\chi^2/N$** | **Runtime (s)** |
| Squared error | $1.90 \times 10^{-3}$ | 25.42 | 32.00 |
| Friedman mean squared error | $1.90 \times 10^{-3}$ | 25.42 | 32.26 |
| **Random Forest Regressor (`random_state=42`)** | **rms** | **$\chi^2/N$** | **Runtime (s)** |
| Squared error (`n_estimators=100`, `min_samples_leaf=6`) | $2.07 \times 10^{-3}$ | 30.10 | 57.61 |
| Friedman mean squared error (`n_estimators=100`, `min_samples_leaf=6`) | $2.07 \times 10^{-3}$ | 30.10 | 62.06 |
| Squared error (`n_estimators=6`, `min_samples_leaf=1`) | $1.47 \times 10^{-3}$ | 15.17 | 34.19 |
| Friedman mean squared error (`n_estimators=6`, `min_samples_leaf=1`) | $1.47 \times 10^{-3}$ | 15.16 | 34.10 |

**Note.** The total number of pixels in our data mask is 104,176. The corresponding figures are presented in the Appendix. There is some natural variation in the runtime for different realizations of the same analysis. The $\chi^2/N$ values are relatively large due to the combination of a high signal-to-noise ratio of the F814W data (S/N = 99.70) and limitations in the cluster mass model, which is fixed in this paper to match the model used in Paper I.

### 3.2.1. RBF Interpolation

RBF interpolation uses the input data as control points to form a set of curved spline functions that provide a smoothed fit to the overall structure. We test the default thin plate spline mode as well as linear and cubic modes to test how different levels of complexity and smoothing fit the compact versus global regions without over-smoothing them, as is often the issue in noise handling.

We implement this using the Python function `scipy.interpolate.RBFInterpolator` (P. Virtanen et al. 2020). Our results are shown in Figure 7 in the Appendix and reported in Table 1. We use an input neighbors number of 6, and we choose a smoothing parameter of 0.25, which we found to have the best balance of not overfitting noise but still providing smoothing. We find that the Linear option performs best for the more compact regions, corresponding to star-forming clumps, to give us smaller residuals there; this is possibly because of how the noise ends up being handled better globally with the lower degree RBF. Each option has similar residuals in the inner caustic regions as seen before in the first test of the other methods. We chose this linear option as the best version of this function, especially since the rms in the image plane residuals are lower for this setup.

### 3.2.2. KNN Regressor

The KNN regression method looks at a chosen number of input pixels around each data point and uses the average of those surrounding pixels to predict the value of the desired point. By design, explicitly utilizing the surrounding region of a given number of points in our scattered, delensed data, this works to prevent excessively large gradients or breaks in the data along that varying resolution.

We implement this using the Python function `sklearn.neighbors.KNeighborsRegressor` (F. Pedregosa et al. 2011). Our results comparing data organization options are shown in Figure 8 in the Appendix and reported in Table 1. This method has two main options to organize the data: (1) ball tree, which recursively partitions data regions into hyperspheres centered around pivotal or important points in the data, and (2) a k-dimensional tree (KD tree), which recursively partitions data regions into k-dimensional space while doing the parameter search. The ball tree method is slightly faster and better suited for organizing large datasets than the KD tree method, which is evident in the smaller rms in the image plane residuals for the former method. Besides speed, the ball tree is also better suited for fitting the smaller-scale data while ensuring smoothness. Considering this as well as its lower rms in the image plane residuals, we therefore choose the KNN ball tree method as the best input option for this function.

This method performs the fastest in our tests. As before, we choose the input number of nearest neighbors as 6 for consistency and not overfitting noise; we additionally test 18 nearest neighbors to check how including more points affects the results for varying resolution regions. For all these tests, in order to ensure smoothing not overfitting noise, we use the uniform-weighted option, as opposed to the distance-weighted.

We next test this option for different numbers of neighbors. We show this result in Figure 9 in the Appendix and report it in Table 1. Lower numbers of neighbors would fit smaller features well, but setting them too low runs the risk of overfitting noise. Higher numbers of neighbors, by contrast, may smooth out the smaller-scale features in the spiral arms of the source galaxy.

Our previously used value of 6 for the number of neighbors is the best suited for ensuring we do not oversmooth the compact features but that we do still perform reasonable noise smoothing.

#### *3.2.3. Decision Tree Regressor*

DTR is a form of organizing the data by splitting it into branches that each have leaf nodes, which each represent a prediction formed from the mean of those values within the leaf. It begins with a higher-level structure and moves down to smaller structures to define. One of the defining options we can adjust and test is the minimum number of samples to form a leaf node, which sets the level of finer detail to include in each prediction; increasing it will consequently smooth out noise.

We implement this using the Python function `sklearn.tree.DecisionTreeRegressor` (F. Pedregosa et al. 2011). Our results are shown in Figure 10 in the Appendix and reported in Table 1. As with the previous function, we choose a single value for the minimum samples to form a leaf as 6, and we test two different criterion options, squared error and Friedman mean squared error, that determine when the function splits the data into branches; the other options did not perform well for our data. We can see that there is more smoothing done in regions outside of where the other functions we tested had smoothing, and this shows up in the overall slightly higher rms in the image plane residuals. These options both perform similarly, so we choose the squared error version for comparison with the other methods.

#### *3.2.4. Random Forest Regressor*

RFR builds on the DTR method by averaging the predictions of an assembled forest of decision trees, each structured as above and with its own full set of predictions. It also uses bootstrapping, which resamples different versions of the data to work against overfitting the statistical noise. The two quantities that we choose to vary are the minimum leaf size, defined above, and the number of estimators, which represents the number of decision trees in the random forest.

We implement this using the Python function `sklearn.ensemble.RandomForestRegressor` (F. Pedregosa et al. 2011). Our results are shown in Figure 11 in the Appendix and reported in Table 1. Similar to before, we test two criteria for the function to split the data into branches: squared error and Friedman mean squared error, testing both 6 as the minimum leaf size (versus the default value of 1), as well as 6 for the number of estimators (versus the default value of 100), to understand their effects on the fit. We see that the models with 6 estimators and the default minimum leaf size perform faster and have lower rms residuals in the image plane as compared to the models with the default number of estimators and a 6 minimum leaf size. We therefore choose the mean squared error criterion with 6 estimators and the default minimum leaf size as our best version of this function, since these settings result in less smoothing in the regions where we expect the source not to need much smoothing.

The benefit of utilizing multiple decision trees is a lower chance of overfitting noise while still maintaining the overall structure of the data. We do indeed see more smoothing, less overfitting of noise, and a more reasonable resolution with the RFR method as compared to the DTR method.

## 4. Results

### *4.1. Initial Results: Mock Data*

We first use our mock arc to test source reconstruction with optimal hyperparameters for each of the four smoothing methods. Figure 2 shows results for the idealized mock arc with no noise, while Figure 3 shows the corresponding results for the noisy mock arc.

In the case of the idealized mock arc, any residuals that cause $\chi^2/N$ to be nonzero arise from smoothing. Each of the smoothing functions performs slightly differently for the more compact regions, though the most smoothing should occur within the caustic region, which is naturally higher resolution from the higher density of the delensed signal and corresponding noise. Looking at the source residuals in Figure 2 and their $\chi^2/N$ values, we see that the KNN and DTR methods perform more smoothing than the RBF and RFR methods but do so in different regions. We note that KNN focuses more on the densely populated region with more differences residing within the caustics, while DTR focuses more on smoothing the highest signal-to-noise star-forming clumps on the lower left of the image. The two star-forming clumps were similarly over-smoothed in our previous `pixsrc` source reconstructions; it is important to narrow down our preferred smoothing method to something that smooths large differences in gradients between adjacent pixels but avoids over-smoothing regions of interest in the case of source studies or transients. Overall, all four methods produce reasonable results.

With the noisy mock arc, the residuals are dominated by noise, and we can see in Figure 3 how each of the smoothing functions handles the noise in the galaxy versus the background regions. In this case, we expect $\chi^2/N \approx 1$ unless the method overfits noise, in which case $\chi^2/N$ will approach zero. We see that lower $\chi^2/N$ values again correspond to the RBF ($\chi^2/N = 0.39$) and RFR ($\chi^2/N = 0.27$) methods, which can be seen in the source residuals as well, though the RBF method performs more smoothing within the denser caustic region. The KNN ($\chi^2/N = 0.87$) and DTR ($\chi^2/N = 0.67$) both perform more smoothing overall, uniformly both in the galaxy signal region and on the background; we also note that the source in the KNN test is visually smoother overall compared to the DTR. We interpret this as a slight overfitting of the noise in the tests with lower $\chi^2/N$, but the similar rms and run times do not indicate that we should have a strong preference for which method. It is important to balance not overfitting noise with not introducing predicted artifacts in the model arc.

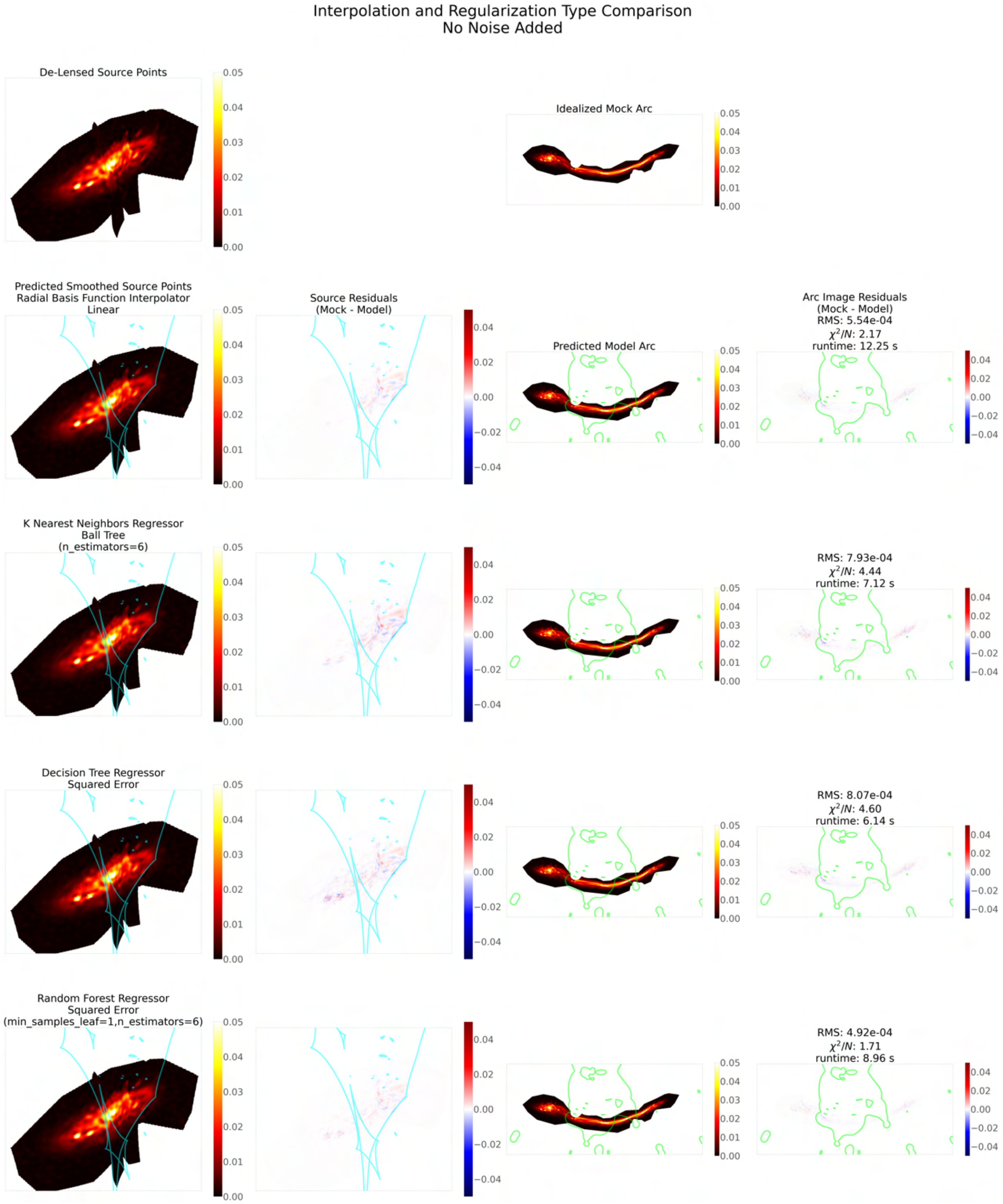

**Figure 2.** Results from our delensing source reconstruction applied to the idealized mock arc with no noise and no PSF. Top row: delensed source prior to smoothing and isolated mock arc. Remaining rows: smoothing method tests on the same set of pixels as above: RBF interpolation, KNN regression, DTR, and RFR. First column: model source, with lensing caustics overlaid for reference. Second column: residuals between the model source and the delensed input points. Caustics overlaid for reference. Third column: model arc created from the smoothed source. Critical curves overlaid for reference. Fourth column: residuals between the model arc and the mock data within the data mask region. In order to consistently compare with the rest of our analysis, we use the F814W data noise level of 0.00038 to compute $\chi^2/N$. Critical curves overlaid for reference. Colorbar units are in electrons per second throughout this article.

### 4.2. Final Choice of Method

We choose the inputs from each category with the best balance of source smoothness and image plane residual rms to test on the data arc again in our comparison Figure 4 and Table 1. We report both the rms and $\chi^2$ for reference, but we use the rms to determine our choice of adaptive smoothing method because it is less dependent on data mask size. All of the methods and options we chose in the aforementioned Python functions were chosen to have interpolation times on the order of a few seconds for our 104,176 pixel, data masked input arc. From Figure 4, we can see which methods overfit the noisy detail from the more densely sampled areas, the delensed points at and within the caustics, and are unable to smooth them as well as other methods; this ranking aligns with the calculated rms and $\chi^2/N$ values in Table 1.

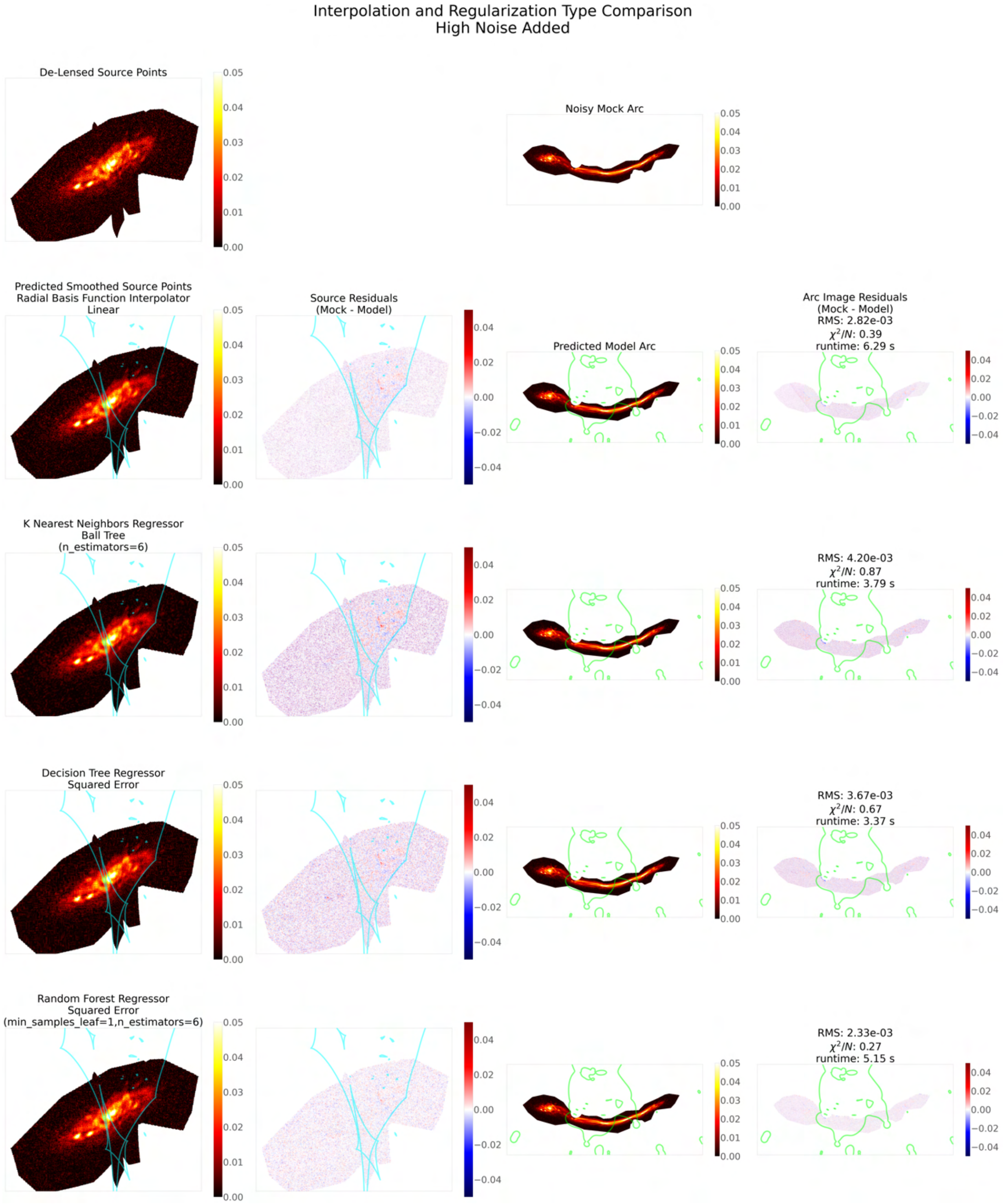

**Figure 3.** Similar to Figure 2 but for the noisy mock arc with no PSF.

Ideally, our methodology should fit the giant arc signal well and smooth the random noise present in the entire image. We note that each of the adaptive smoothing methods we tested appears to fill out the upper right section of the source, the region for which we have fewer delensed points from the giant arc. In each case, we can see that the area with the highest smoothing is located in the central and rightmost parts of the source, the region with the most delensed points and highest magnification that resides within the caustics our previously optimized model predicted. We additionally expect some of the residuals to be the result of a slightly incorrect cluster mass model, which can be further optimized.

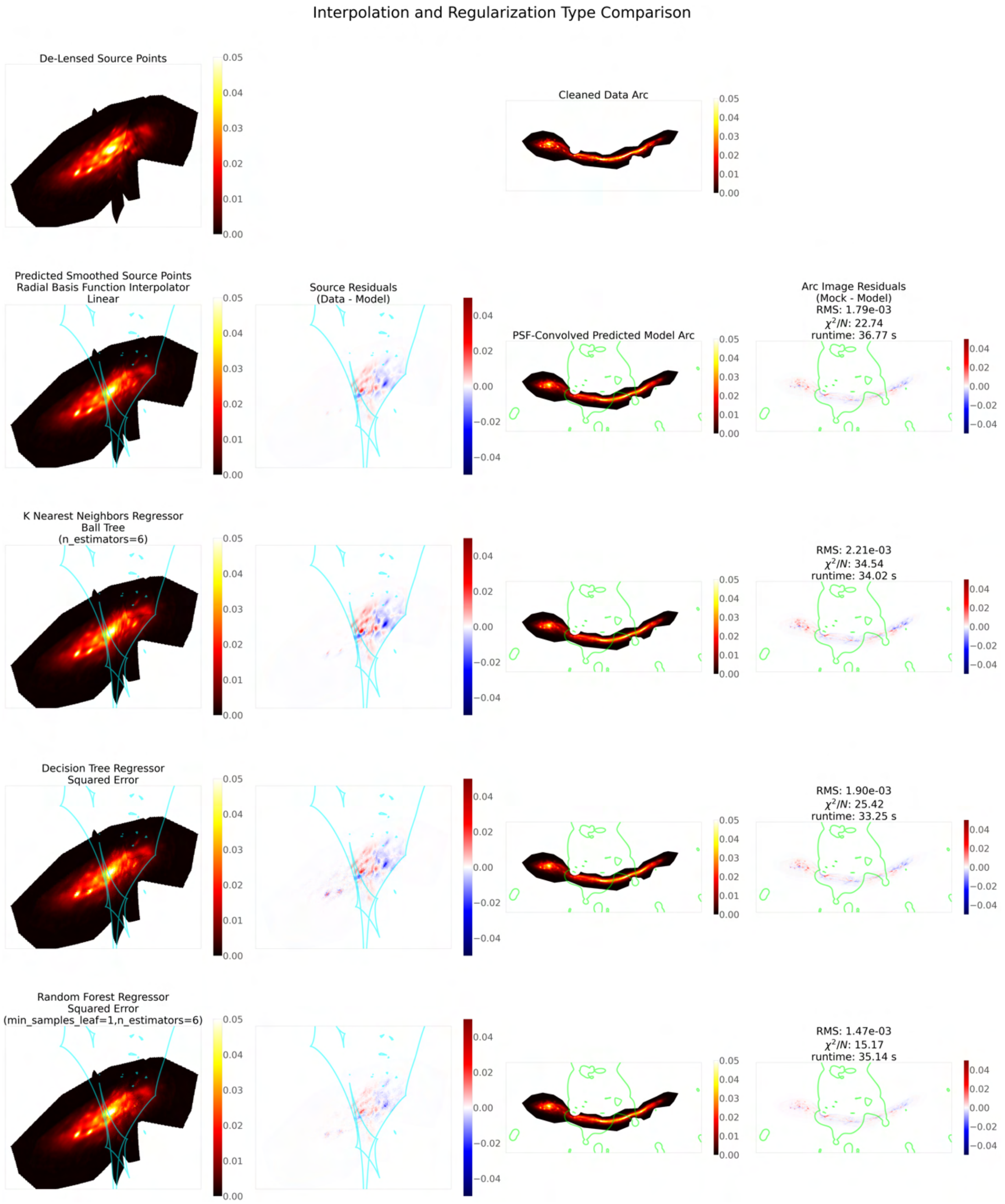

**Figure 4.** Results from our delensing source reconstructions with the same structure as Figure 2, now applied to the isolated HFF giant arc data in the F814W filter, with the additional step of deconvolving and reconvolving the HFF F814W filter data with the corresponding PSF. Shown here is a set of the best input options from each of the source plane adaptive smoothing methods: RBF interpolation, KNN regression, DTR, and RFR.

Based on a combination of visual inspection for unphysical smoothing artifacts, low rms residuals, uniform smoothness, corresponding $\chi^2/N$, and a reasonably short runtime, the KNN routine with the ball tree option and an appropriate number of nearest neighbors for this data offers the most reasonably smoothed and reconstructed source. We note that the ideal number of neighbors value will be different for each dataset based on resolution and amount of noise; we will further investigate how to best determine this based on resolution in our follow-up study when we constrain our cluster model with this source reconstruction code. Ensuring the smoothing method takes a short amount of time is critical in order to use the delensing code as a constraint in a parameter optimization search for cluster modeling.

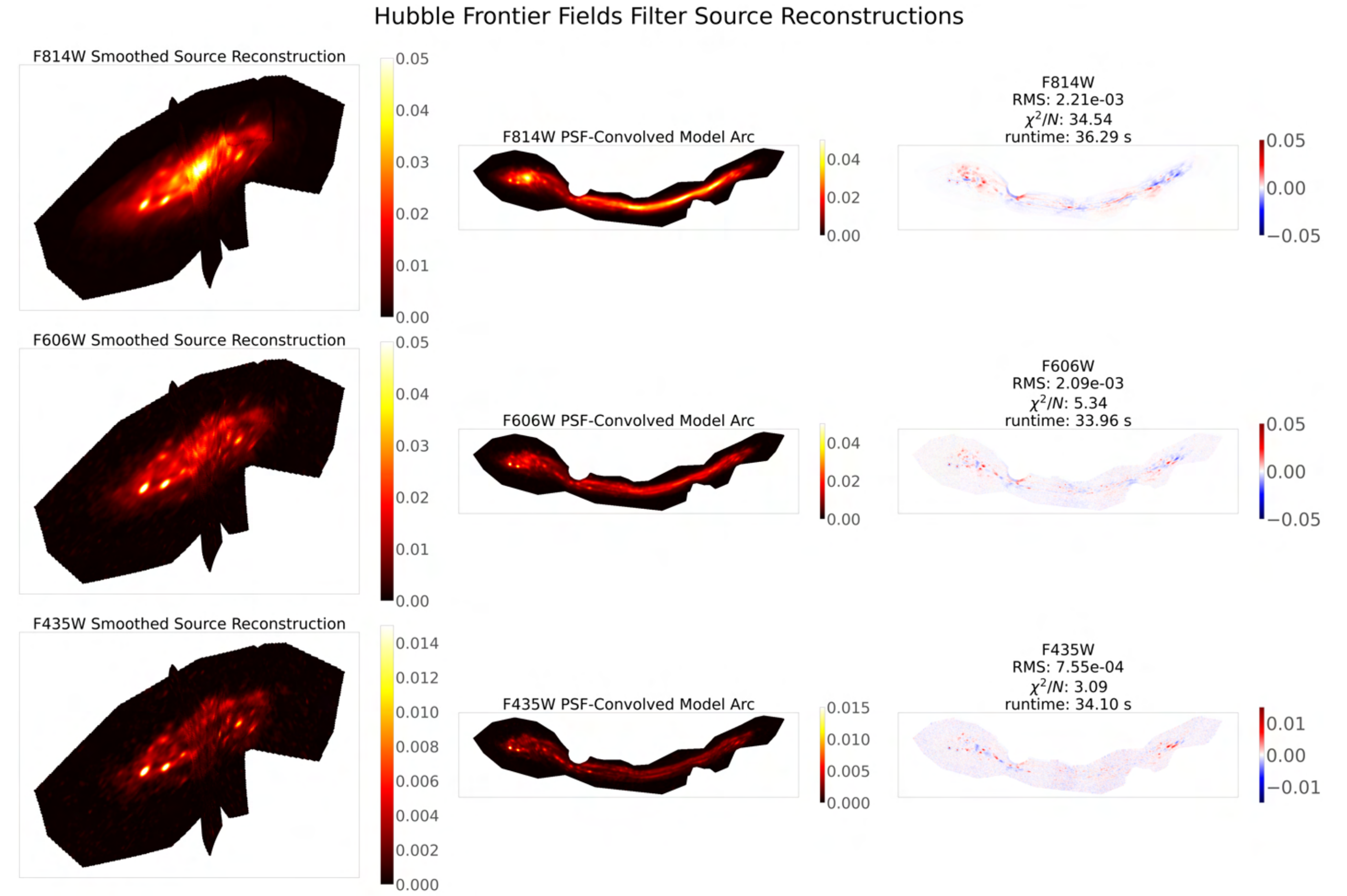

**Figure 5.** A370 data arc source reconstruction for the F814W, F606W, and F435W filters, predicted and smoothed with the KNN ball tree option with 6 nearest neighbors. We display here the source reconstruction in the left column, the reconvolved model arc in the middle column, and the residuals showing deconvolved data minus model arc with no PSF reconvolution in the right column.

## 5. Discussion

In order to produce a comparison to our RGB image in Paper I, we perform the delensing source reconstruction on the F814W, F606W, and F435W filters from the HFF with the same cluster model optimized in Paper I. The red filter has the least noise to handle, but the green and blue filters have more significant structure in their noise, which was more difficult to smooth out. The blue filter especially had noise levels comparable to the signal.

We use the same input settings of 6 nearest neighbors for the KNN ball tree option to obtain the source reconstructions for each filter, shown separately in Figure 5 and assembled as RGB images in Figure 6.

This comparison figure to the results from Paper I shows that the source reconstruction performs similarly well and has a similar structure in the regions that can still be optimized further in the cluster model, such as the caustic boundaries and the smaller-scale structure. We compare the rms in the residuals in Table 2. We recomputed the rms and $\chi^2/N$ values strictly within the displayed data mask for side-by-side comparison between our methodology here and our `pixsrc` results from Paper I. We can see that the two source reconstruction methods produce results of similar statistical quality, while `SourceREACH` is much faster.

Looking at the source residuals columns in the simplified models in Figures 2 and 3, we can see that the source smoothing happens in regions with compact brightness sources, such as star-forming clumps, and especially within the caustic region. This is even more apparent in the source residuals columns in the full data from Figure 4. The model arc with added Gaussian noise showed more differences between how each function smoothed the noise in the background areas and showed a similar structure in the residuals.

Considering the arc residuals in corresponding figures, we see that there is some structure in the residuals near the galaxies that we were previously (Paper I) unable to remove and simply masked for our source reconstructions. This can further be improved in the future by more accurately modeling and masking the galaxies, perhaps with higher-resolution data as well. The residuals show a similar pattern to what we saw in Paper I, where the F814W source reconstruction fits the larger-scale surface brightness regions better, and the F814W and, more so, the F435W filter source reconstructions have structure in the residuals near the star-forming clumps represented by more compact brightness regions. The distribution of the brightness peaks and extended structure in each filter varies, as each originates from a different region, such as the center of the spiral galaxy being redder and more extended, the star-forming clumps being bluer and more concentrated, and the green filter being an intermediate example between the two. Since this is all done using our locally optimized model from Paper I, we anticipate that once

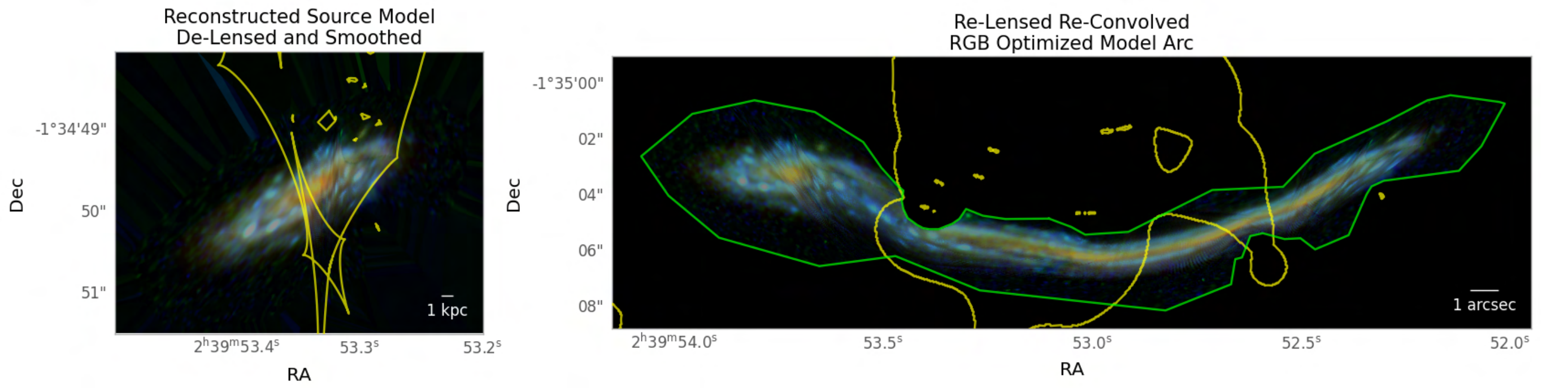

**Figure 6.** Left: RGB source reconstruction using the F814W, F606W, and F435W filters. Corresponding caustics are overlaid in yellow; a 1 kpc scale bar is shown for reference. Right: model arc created using source reconstruction and reconvolved with PSF in each filter. Corresponding critical curves and data mask are overlaid in yellow and green, respectively; a 1″ scale bar is shown for reference.

**Table 2**
Comparison of Rms Values for F814W, F606W, and F435W Filters between the Previous `pixsrc` Analysis and the New `SourceREACH` Analysis

| HST Filter from HFF | S/N for Arc | rms `pixsrc` | rms `SourceREACH` | $\chi^2/N$ `pixsrc` | $\chi^2/N$ `SourceREACH` |
|---|---|---|---|---|---|
| F814W | 99.70 | $2.189 \times 10^{-3}$ | $2.186 \times 10^{-3}$ | 33.82 | 34.54 |
| F606W | 51.01 | $2.073 \times 10^{-3}$ | $2.691 \times 10^{-3}$ | 5.27 | 5.34 |
| F435W | 65.52 | $8.436 \times 10^{-4}$ | $7.563 \times 10^{-4}$ | 3.86 | 3.09 |

**Note.** Note that the two methods are comparable statistically, though SourceREACH is able to run much faster. The signal-to-noise ratio is higher for the F814W filter than for the F606W and F435W filters, leading to different $\chi^2/N$ values. We also indicate the measured signal-to-noise value in the giant arc region for each filter.

we include a $\chi^2$ from the source reconstruction in one or all filters as a constraint in the cluster model along with the point-image $\chi^2$ fitting, we may be able to improve these residuals and gain an even more accurate source reconstruction and cluster model, especially for this region, while affecting the fit of the global point images even less than we previously did with our optimization.

## 6. Conclusions and Future Work

We have introduced a new source reconstruction methodology called SourceREACH that can efficiently delens and interpolate giant arcs lensed by galaxy clusters. This method uses the full, pixelized giant arc information in a short interval of time, which is crucial for folding in the extended image fit as a constraint in cluster modeling.

Our delensing methodology works with solely the input set of image points; we therefore avoid the need to construct and manipulate large matrices that would otherwise limit how we use source reconstruction as a cluster modeling constraint. In order to speed up the source reconstruction process, we tested techniques from big data analysis to adaptively smooth our two-dimensional source model.

The smoothing method, whether interpolation or regression, must be chosen carefully, as something that has too-small a residual rms will actually overfit the delensed noise, and some options may end up smoothing out features of interest such as star-forming clumps or transient events. Our choice ends up being a close decision between several functions that performed very similarly, but the identification of the KNN ball tree technique as the ideal method was informed by testing each on multiple filters and quantifying where and why certain issues occurred in each fit. We find that the smoothing in the KNN produces the most reasonable source; it is strong enough to visibly smooth the noise in the source plane in a more uniform way than in the other tests, and it avoids over-smoothing specific regions that would be of interest to source or transient studies.

We will next use `SourceREACH` to update lens models for A370 that simultaneously incorporate constraints from the giant arc and the abundance of compact lensed images. Our methodology could also be valuable for upcoming large surveys that include cluster lenses, such as Euclid (Euclid Collaboration et al. 2025) and LSST (LSST Science Collaboration et al. 2009; A. J. Shajib et al. 2025), where the data may not be as deep as for the HFF (J. M. Lotz et al. 2017) or CANUCS and other surveys with JWST (J. M. Diego et al. 2024, 2025; G. T. E. Sarrouh et al. 2026) but may contain interesting systems that would need in-depth, efficient analysis to determine when transient events (e.g., J. M. Diego et al. 2024; B. L. Frye et al. 2024; A. Agrawal et al. 2026) may be observed in follow-up studies without needing separate, higher-resolution observations first. We will investigate this in our future work.

Overall, it is important not to ignore the full information present in high-resolution, extended giant arcs lensed by galaxy clusters, as they provide information that cannot be assumed from point-image fitting in a straightforward manner. The delensing methodology presented here is a step forward in ensuring we can easily utilize the source reconstruction of these critical curve-crossing lensed images and fully describe the cluster region causing these maximally magnified images, as well as the source itself, in accurate detail.

## Acknowledgments

We acknowledge financial support from the Rutgers–New Brunswick Department of Physics and Astronomy. We thank José María Diego, James Nightingale, Giovanni Granata, and Huimin Qu for helpful discussions. We also thank the anonymous referee for the constructive feedback on the original manuscript.

## Data Availability

The data, mock data, previously optimized model, and Jupyter notebook source reconstruction code used in this paper are available at https://github.com/LanaEid/SourceREACH. The original data for the HFF can be accessed at https://archive.stsci.edu/prepds/frontier/.

## Appendix
## Adaptive Smoothing Method Tests

We present here the full set of tests we performed in order to determine the best parameters for the different adaptive smoothing methods, each described in Section 4 and implemented in Python: RBF interpolation in Figure 7, KNN regression in Figures 8 and 9, DTR in Figure 10, and RFR in Figure 11.

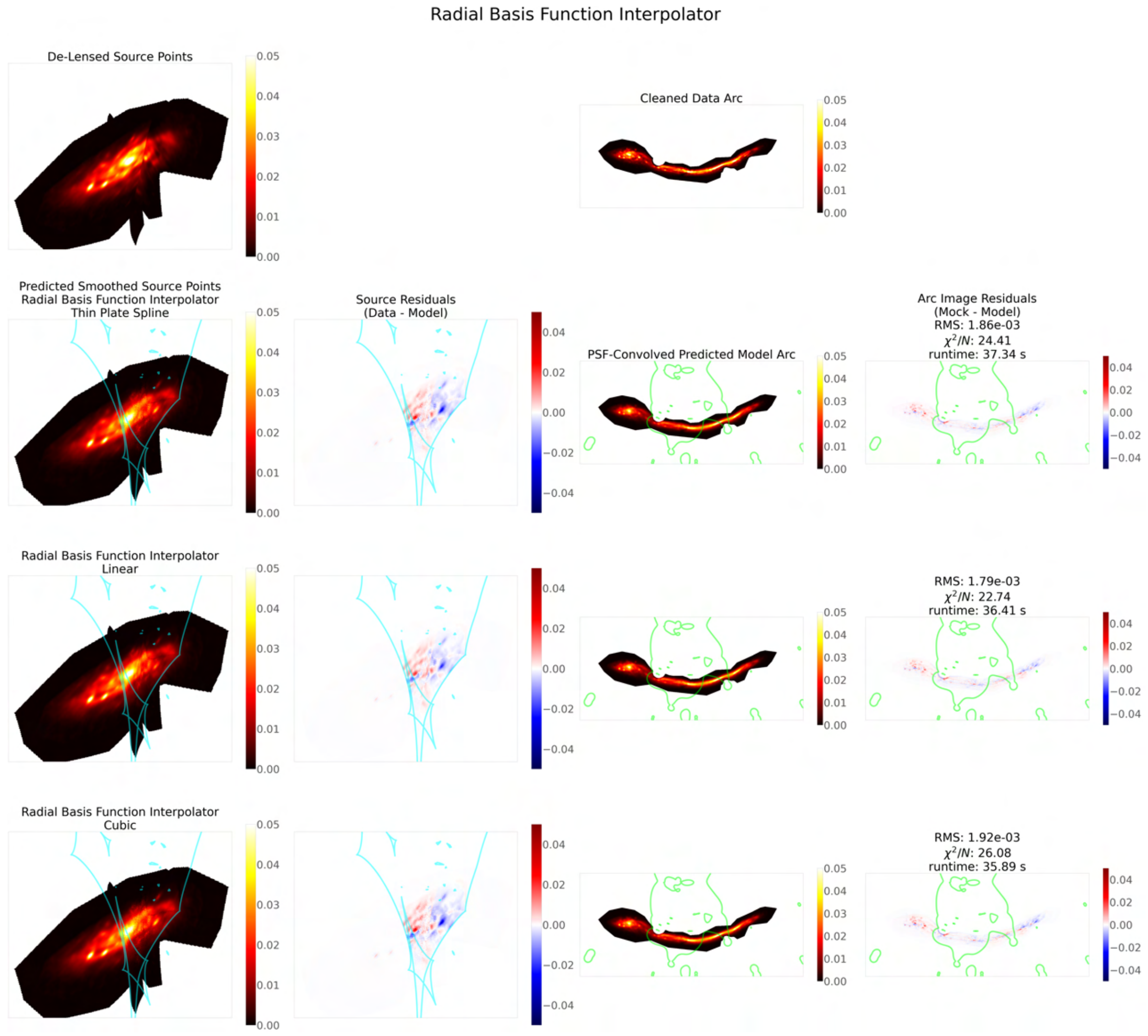

**Figure 7.** Results from our delensing source reconstruction applied to the cleaned data in the F814W filter for the observed arc in A370. The structure of the figure is the same as in Figure 2; note that here we compare the PSF-convolved model arc to the cleaned data arc. Here, we focus on RBF interpolation with different options. This interpolation algorithm constructs functions defined as the distance between inputs and fixed centers and interpolates using their weighted sum. The RBF can be chosen as a polynomial of the desired degree for different levels of detail.

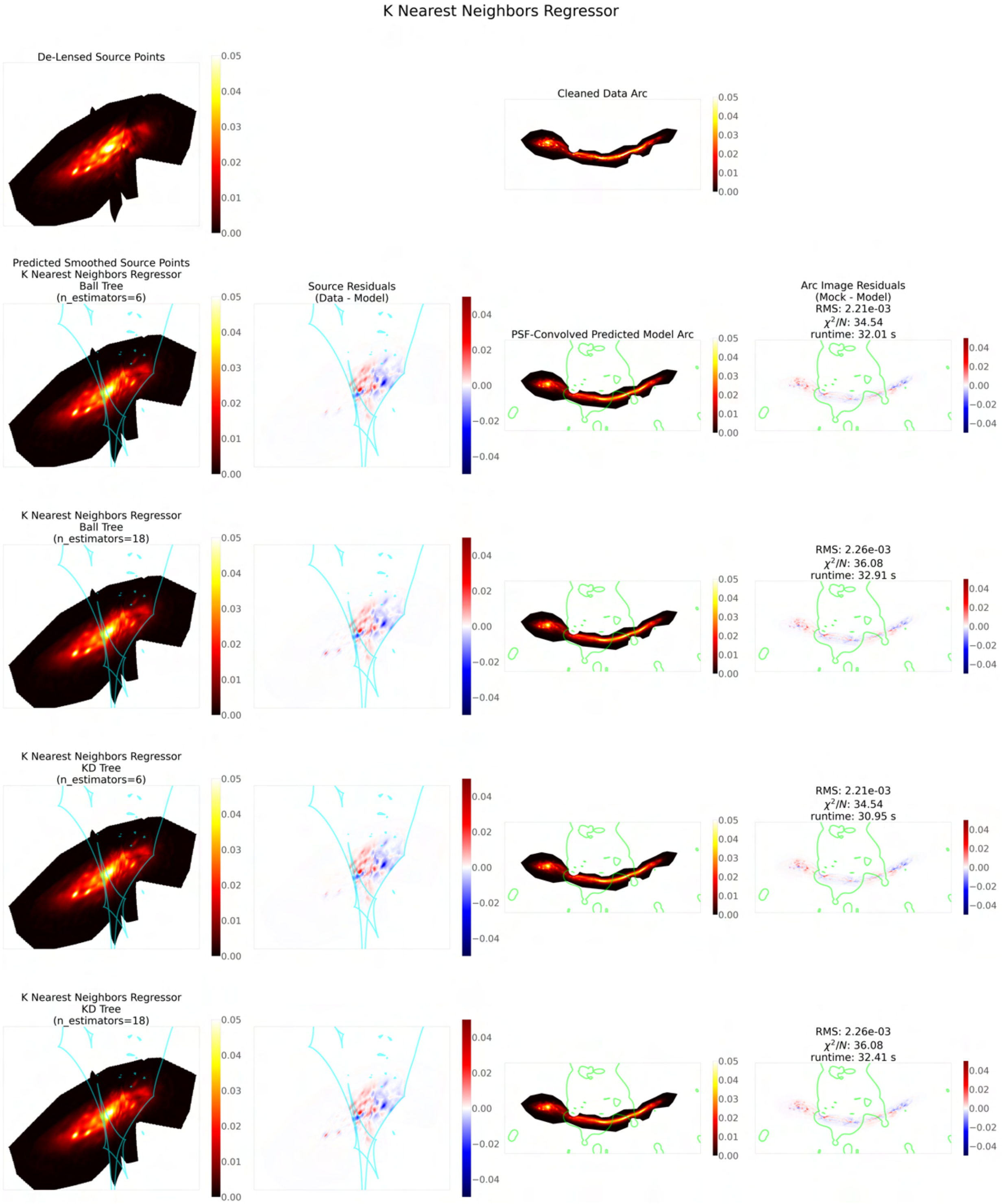

**Figure 8.** Similar to Figure 7 but for different organization routine options for KNN regression. This supervised machine learning algorithm predicts the value at a given point by averaging the values of the KNN.

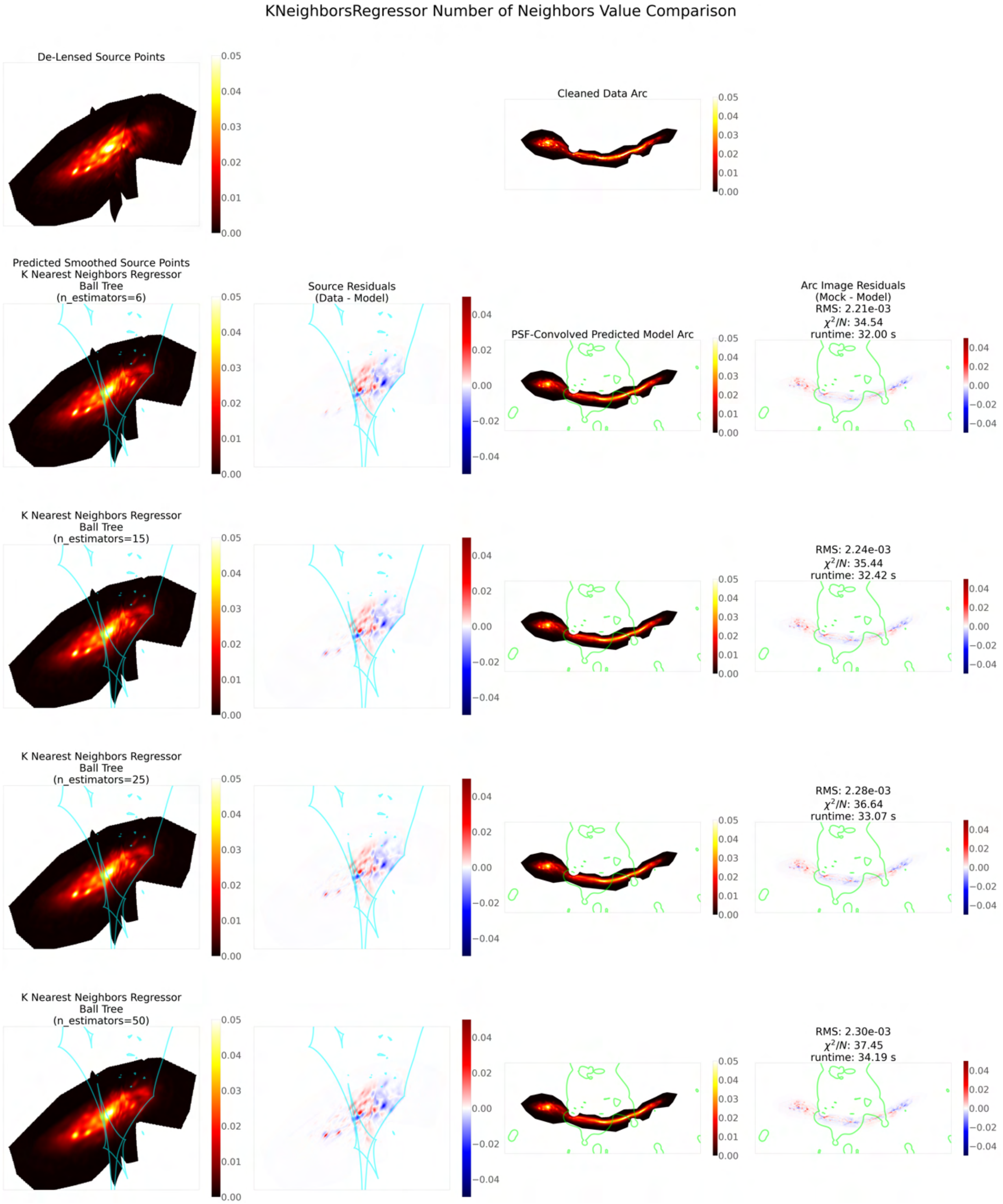

**Figure 9.** Similar to Figure 7 but for the KNN ball tree smoothing option with a varying number of neighbors. The best option for dealing with the scattered delensed noise was 6 nearest neighbors.

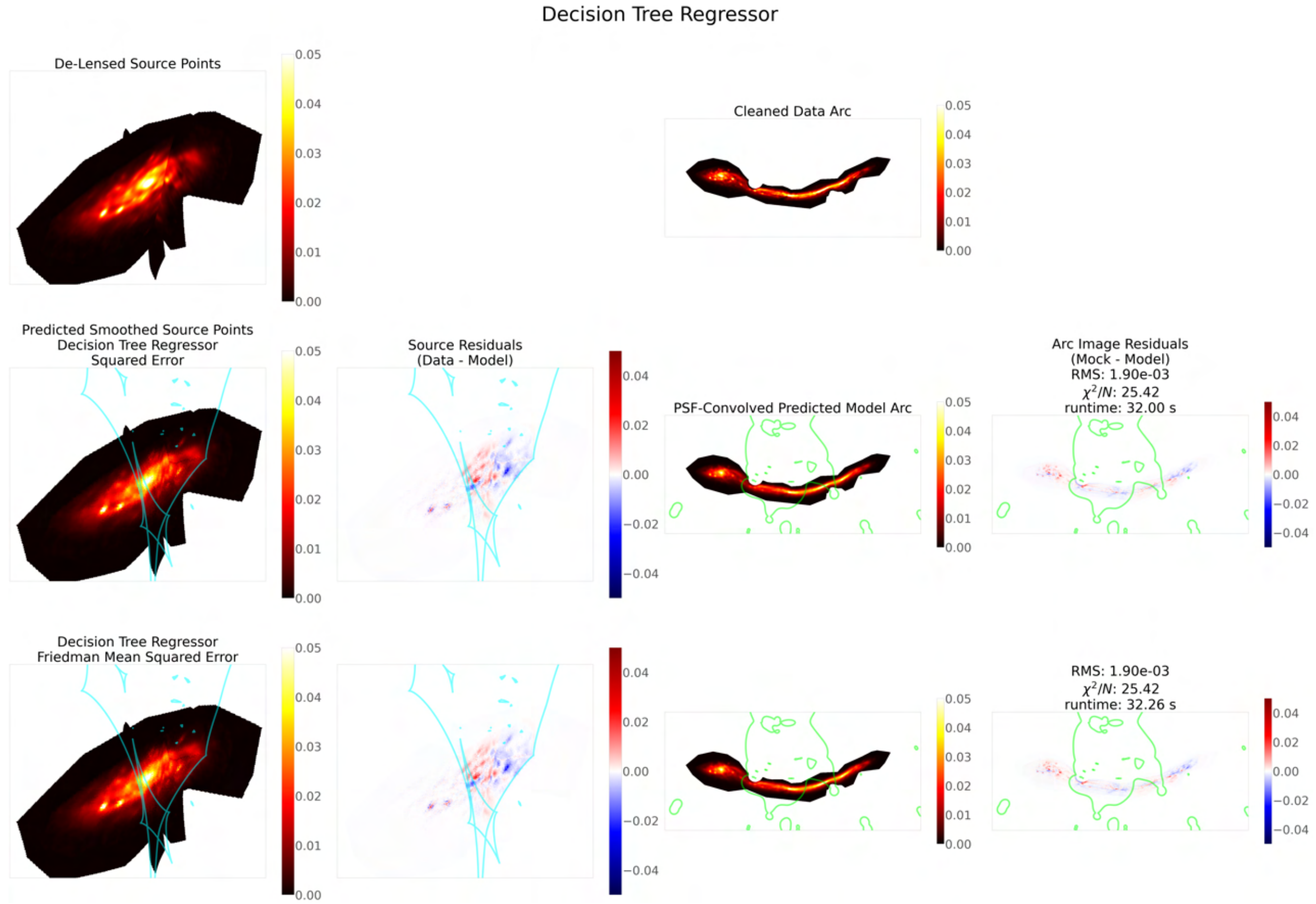

**Figure 10.** Similar to Figure 7 but for different interpolation routine options for the DTR method. This supervised machine learning algorithm recursively partitions the input data to fit smaller sections. It minimizes the error by identifying a branching threshold at which to form leaves that it averages to fit each value.

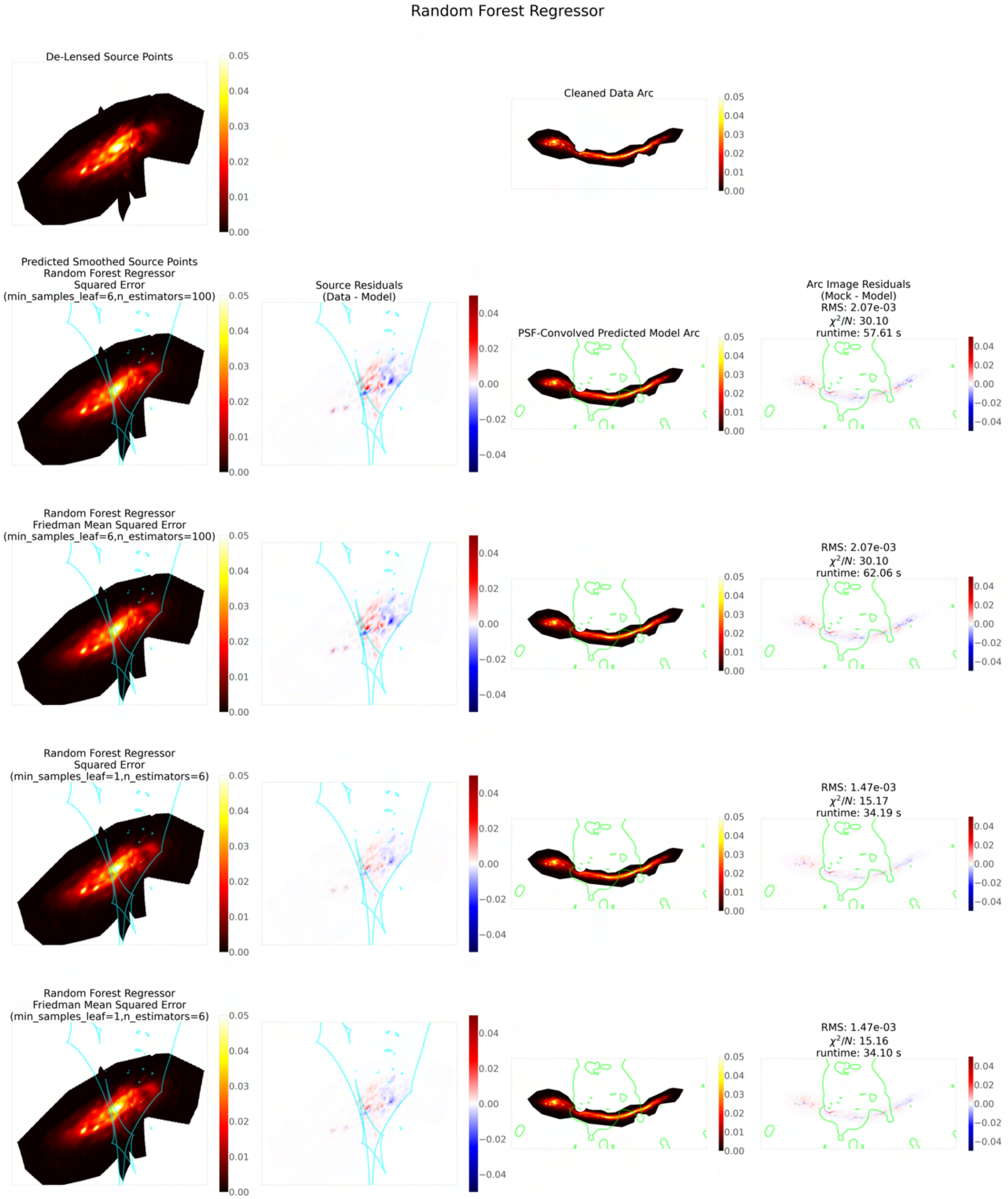

**Figure 11.** Similar to Figure 7 but for different interpolation routine options for the RFR method. This ensemble machine learning algorithm averages conclusions from multiple decision trees to avoid overfitting. Each tree is trained on a randomly selected and bootstrapped section of the data, and it recursively splits the inputs at a given threshold to minimize error and predict values.

## ORCID iDs

Lana Eid https://orcid.org/0009-0003-5696-9355
Charles R. Keeton https://orcid.org/0000-0001-6812-2467